# A novel TSK fuzzy system incorporating multi-view collaborative transfer learning for personalized epileptic EEG detection

Andong Li [a, b], Zhaohong Deng [a] *, Qiongdan Lou [a, c]

[a]School of Artificial Intelligence and Computer Science, Jiangnan University and Engineering Research Center of Intelligent Technology for Healthcare, Ministry of Education, Wuxi, 214122, China

[b]School of Computer and Information Engineering, Institute for Artificial Intelligence, Shanghai Polytechnic University, Shanghai, 201209, China

[c] School of Internet of Things Engineering, Wuxi University, Wuxi, 214122, China

**Abstract—**Electroencephalography (EEG) plays a critical role in the diagnosis of epilepsy; however, personalized EEG-based seizure detection, in which models are trained for specific individuals, still has several challenges, including limited feature selection, insufficient training data, and inconsistent cross-domain distributions. To address these problems, we propose a Takagi-Sugeno-Kang fuzzy system (TSK FS)-based epilepsy detection algorithm that integrates multi-view collaborative transfer learning (MVTL-FS). The proposed method increases feature diversity through multi-view feature extraction, mitigates data scarcity in the target domain via source-domain knowledge transfer, and achieves cross-domain distribution alignment by incorporating the maximum mean discrepancy criterion. By integrating transfer learning with multi-view modeling within a unified framework, the approach effectively improves the generalization capability of personalized models. More importantly, the rule base of the TSK fuzzy system not only provides powerful nonlinear modeling capability but also increases model transparency and robustness through its fuzzy inference mechanism. On the CHB-MIT dataset, 20 transfer learning tasks were performed, and the proposed method achieved an average accuracy of 97.32%, with improvements of 2.39% over the best non-multi-view baseline and 2.79% over the best multi-view baseline. Moreover, model analysis further validated the robustness of the method, revealing its potential for reliable personalized seizure detection.

## 1. Introduction

Epilepsy is a chronic brain disease. The World Health Organization (WHO) estimates that approximately 50 million people have epilepsy worldwide [1]. The excessive discharge of brain neurons during epileptic seizures, which can be attributed to various factors, causes the central nervous system to malfunction. Because seizures typically occur at night, diagnosing them afterward by relying only on the recall of the past events of the patient is inaccurate. Common diagnostic methods for detecting epilepsy include medical history, physical examination, and auxiliary examinations such as electroencephalography (EEG), which are important clinical examination methods. EEG is used to measure the electrical activity of brain cells [2, 3]. Observations from EEG activities can be used effectively for clinical diagnosis. In one study, the number of seizures detected by EEG was 29 times and 7 times greater than the number of episodes detected by family members and nurses, respectively [4].

In addition to epilepsy, EEG is also essential for the diagnosis of various neurological disorders [5-8]. For example, Xi et al. analyzed the topological properties of EEG-based functional brain networks in stroke patients and reported that stroke reduces network complexity and connectivity [7]. In another research direction, EEG has shown great potential for Alzheimer's disease identification; the N-TSK method, which integrates functional network features with fuzzy learning, achieved superior accuracy and interpretability, creating new avenues for automated neurological diagnosis [8]. These advancements highlight the expanding role of EEG-based intelligent analysis in neurological research, of which epilepsy detection has been among the earliest and most extensively explored applications.

In particular, since Gotman [9] first proposed the widely used epilepsy detection method, the application of intelligent algorithms to assist clinicians in epilepsy detection has received increasing attention [10]. The process of automatic epilepsy detection with EEG signals can be broadly divided into three steps: EEG signal acquisition, discriminative feature extraction, and classifier construction. In this paper, we focus on the latter two steps, namely feature extraction and classifier design.

Feature extraction constitutes a crucial step in obtaining meaningful information from EEG signals and serves as a basis for effective model training. Different feature extraction methods can identify different types of features, such as time domain, frequency domain, time-frequency domain, and non-linear features [11-13]. Each category of features has its own advantages: time domain features are simple and intuitive; frequency domain features reveal variations in the frequency components of EEG signals; time–frequency features jointly characterize temporal and spectral information; and non-linear features (such as fractal dimension, entropy, and fuzzy entropy)

* Corresponding author at: School of Artificial Intelligence and Computer Science, Jiangnan University and Engineering Research Center of Intelligent Technology for Healthcare, Ministry of Education, Wuxi 214122, China.

*E-mail address*: dengzhaohong@jiangnan.edu.cn

measure signal complexity and mitigate the effects of non-stationarity. However, different feature types also have their own limitations. Therefore, selecting the appropriate feature extraction method on the basis of specific application scenarios or data characteristics remains challenging.

In the construction of classifiers, EEG signal recognition is typically formulated as either a binary or a three-class classification problem. In the binary setting, the objective is to distinguish seizure states from normal states. In contrast, the three-class formulation further categorizes EEG signals into normal, interictal, and ictal states. Neural networks [14, 15], naive Bayes [16], support vectors [17], and fuzzy systems (FS) [18, 19] are typically used in binary classification. Recurrent neural networks [20], support vector machine (SVM) [21], K-nearest neighbor (KNN) [22], decision trees with the C4.5 algorithm [23], and FSs [24] have been used for three-class classification of EEG signals. In addition, end-to-end deep learning methods have been developed for EEG analysis in recent years [25, 26]. Among the abovementioned methods, fuzzy rules and fuzzy inference-based FSs are distinctive in their capability to maintain a good balance between accuracy and interpretability.

However, challenges remain despite the advances in automated epilepsy detection [27, 28]. The difficulty of extracting targeted discriminatory features for identification is an important issue. In addition to conventional approaches that extract features from a single view, multi-view feature extraction and learning techniques have been used to address this challenge, which can effectively exploit the complementary information of different views to improve the detection performance [29]. Another challenge is the lack of sufficient samples for training effective patient-specific models. Transfer learning techniques have been used here because of their ability to improve the learning of specific tasks on the basis of the existing knowledge in the related tasks [30, 31]. While these two challenges are alleviated via multi-view learning and transfer learning, they are usually not addressed simultaneously and deserve in-depth research.

To address these challenges, we propose a novel multi-view transfer learning TSK fuzzy system (MVTL-FS) for epilepsy EEG detection. This framework combines the advantages of multi-view learning and transfer learning within a unified architecture and uses the interpretability and fuzzy inference capabilities of rule-based models to improve both classification performance and increase model transparency. MVTL-FS integrates features from multiple views derived through diverse methodologies, enabling complementary information utilization across views. Specifically, in this paper, we integrate features from the time domain view, frequency domain view, and time-frequency view. This multi-view integration increases the classification accuracy while reducing the dependence on any single view. Furthermore, transfer learning is employed to integrate knowledge from a source domain into the target domain, effectively mitigating the issue of data scarcity in patient-specific epilepsy detection. The fuzzy system increases the robustness of the model by effectively handling uncertainty through the application of fuzzy rules. Additionally, MVTL-FS improves model interpretability by elucidating the logical relationships between input features and outputs. Finally, a multi-view TSK FS classifier is constructed.

The main contributions of this work are summarized as follows:

1) By integrating multi-view learning with transfer learning, a collaborative multi-view transfer learning method i.e., MVTL-FS, is proposed for classification model construction based on the TSK FS. Multi-view learning enables the proposed model to incorporate between-view knowledge and reduces the difficulty in selecting a specific method that is appropriate for feature extraction. Moreover, transfer learning enables the model to have robust performance in complex scenarios.

2) For patient-specific epileptic EEG detection, we propose an epilepsy detection framework based on MVTL-FS, named TSK-MVTL-FS. This framework provides detailed guidance for utilizing the proposed MVTL-FS for epilepsy detection. Our approach provides greater robustness and transparency than compared to black-box model-based frameworks do.

3) Extensive experimental studies are conducted on the CHB-MIT dataset to demonstrate the effectiveness of the proposed method for patient-specific epilepsy detection.

The remainder of this paper is organized as follows: Section 2 presents the main technical background. The proposed MVTL-FS method is described in detail in Section 3. Section 4 discusses the procedure of epilepsy detection via MVTL-FS. Experimental validation and analysis are reported in Section 5. Conclusions and future works are given in Section 6.

## 2. Related work

### *2.1 Multi-view learning for epilepsy detection*

Multi-view learning is a learning paradigm that aims to enhance modeling performance by leveraging data representations obtained from multiple views [32-35]. It addresses the inherent limitations of single-view approaches by promoting feature diversity and integrating complementary information, thereby improving overall performance. The existing multi-view learning algorithms can be classified into three main categories, i.e., co-training [33], multiple kernel learning [34], and subspace learning[35].

Recently, multi-view learning has been applied for epilepsy detection. Tang et al. [36] proposed a multi-view convolutional gated recurrent network framework to analyze the spatiotemporal sequences of multi-view features to identify potential changes prior to seizures. Yuan et al. [37] proposed a unified multi-view deep learning framework based on multichannel scalp EEG signals to detect abnormalities associated with epileptic seizures. Liu et al. [38] used a multi-view convolutional neural network framework to predict the occurrence of seizures. Tian et al. [39] proposed a multi-view deep feature extraction method combined with an interpretable classifier for epilepsy detection.

### *2.2 Transfer learning for epilepsy detection*

Transfer learning refers to the application of knowledge and/or data from a domain to a different but related domain. The objective of transfer learning is to identify the similarities between the domains to address challenging data mining problems such as big data, unlabeled learning, universal models, or personalized applications [40-44]. Transfer learning and related techniques have already been used in many areas, such as WIFI signal localization and image classification [45, 46].

In epilepsy detection, constructing fully universal models is challenging due to the substantial inter-individual variability in EEG signals. Transfer learning offers a potential solution to this challenge by enabling knowledge transfer across subjects, thereby mitigating the limitations of subject-specific models. S. Raghu et al. [41] classified seven different types of seizure versus non-epileptic EEG signals by applying convolutional neural networks(CNN) and transfer learning. Rodrigues et al. [42] proposed a transfer learning method for processing the statistical variability of the EEG signals of different subjects. Jiang et al. [43] integrated a label shift vector into a generalized linear model to adjust the weights in the least squares regression classifier. Xia et al. [44] proposed a cross-domain classification model with knowledge utilization maximization (CDC-KUM).

*2.3 TSK FS for epilepsy detection*

The fuzzy set, which is a promotion of the classic crisp set, is proposed to model the vagueness of data in the real world. Complex knowledge can be represented in imprecise terms via fuzzy set theory and fuzzy logic theory, and the intelligent models thus developed, i.e., fuzzy systems, have good interpretability. Since fuzzy systems were first introduced in [47], various extensions have been developed. The TSK fuzzy system is among the most widely applied fuzzy systems, because it preserves the interpretability of fuzzy rules in the antecedent while improving the learning ability through linear functions in the consequent, achieving both flexibility and modeling accuracy, and it has been widely applied in diverse fields such as data mining and intelligent healthcare [48-52]. Depending on whether the membership functions incorporate uncertainty, fuzzy systems can be further categorized into the type 1 system and the type 2 system: the former is widely used because of its structural simplicity and computational efficiency, whereas the latter introduces uncertainty modeling and is more suitable for complex environments [53-57]. In this study, considering that the feature patterns of the research subjects are relatively well defined and the dataset size is limited, we use a type 1 TSK fuzzy system to strike a balance between modeling performance and computational efficiency. The specific definition of the TSK fuzzy system employed in this study is presented below.

TSK FS contains a fuzzy rule base. The $k$th rule can be defined as follows:

$$\begin{aligned}
&\text{IF: } x_1 \in A_1^k \wedge x_2 \in A_2^k \ldots \wedge x_d \in A_d^k, \\
&\text{Then: } f_j^k(\mathbf{x}) = p_{0,j}^k + p_{1,j}^k x_1 + p_{2,j}^k x_2 + \ldots + p_{d,j}^k x_d, \\
&\qquad k = 1, 2, \ldots, K \\
&\qquad j = 1, 2, \ldots, C
\end{aligned} \tag{1}$$

where $K$ represents the total number of rules, $C$ denotes the total number of classes and $d$ indicates the dimensionality of the sample. The input sample is denoted as $\mathbf{x} = (x_1, x_2, \ldots, x_d)$, where $x_i$ represents the $i$th feature of the input vector. $A_i^k$ represents the fuzzy set of the $i$th feature in the $k$th rule, and $f_j^k(\mathbf{x})$ denotes the $j$th output of the $k$th rule. $p_{i,j}^k$ represents the consequent parameter of the $j$th output in the $k$th rule. $\wedge$ denotes a fuzzy conjunction operation. The output of the TSK FS can be obtained on the basis of (2) with specific fuzzy inference operations used:

$$f_j(\mathbf{x}) = \frac{\sum_{k=1}^{K} \mu^k(\mathbf{x}) f_j^k(\mathbf{x})}{\sum_{k'=1}^{K} \mu^{k'}(\mathbf{x})} = \sum_{k=1}^{K} \tilde{\mu}^k(\mathbf{x}) f_j^k(\mathbf{x}) \tag{2}$$

where $\mu^k(\mathbf{x})$ denotes the firing strength of the $k$th rule for the sample input $\mathbf{x}$, which can be normalized to $\tilde{\mu}^k(\mathbf{x})$. They are usually obtained by

$$\mu^k(\mathbf{x}) = \prod_{i=1}^{d} \mu_{A_i^k}(x_i) \tag{3}$$

$$\tilde{\mu}^k(\mathbf{x}) = \mu^k(\mathbf{x}) \Big/ \sum_{k=1}^{K} \mu^k(\mathbf{x}) \tag{4}$$

where $\mu_{A_i^k}(x_i)$ represents the membership degree of the $i$th element $x_i$ of $\mathbf{x}$, which belongs to the fuzzy set $A_i^k$. The Gaussian function is typically used as the membership function [58], i.e.,

$$\mu_{A_i^k}(x_i) = \exp\left( \frac{-\left(x_i - c_i^k\right)^2}{2\delta_i^{2k}} \right) \tag{5}$$

The parameters $c_i^k$ and $\delta_i^k$ in (5) can be obtained via different methods, such as deterministic clustering [59, 60]. If the antecedent parameters of a TSK FS have been determined, for an input vector $\mathbf{x}$, the $j$th output of the model, i.e., $y_{j,pre}$ can be expressed as a linear model in the new feature space as follows:

$$y_{j,pre} = f_j(\mathbf{x}) = \mathbf{p}_{\mathrm{g},j}^{\mathrm{T}} \mathbf{x}_{\mathrm{g}} \tag{6}$$

with the transformations below,

$$\mathbf{x}_{\mathrm{e}} = [1, \mathbf{x}^{\mathrm{T}}]^{\mathrm{T}} \in \mathbf{R}^{(\mathrm{d}+1)\times 1}, \tag{7a}$$

$$\tilde{\mathbf{x}}^{\mathrm{k}} = \tilde{\mu}^k(\mathbf{x}) \mathbf{x}_{\mathrm{e}} \in \mathbf{R}^{(\mathrm{d}+1)\times 1}, \tag{7b}$$

$$\mathbf{x}_{\mathrm{g}} = \mathrm{concat}(\tilde{\mathbf{x}}^1, \tilde{\mathbf{x}}^2, \ldots, \tilde{\mathbf{x}}^K) \in \mathbf{R}^{K(\mathrm{d}+1)\times 1}, \tag{7c}$$

$$\mathbf{p}_j^k = \left[ p_{0,j}^k, p_{1,j}^k, \ldots, p_{d,j}^k \right]^{\mathrm{T}} \in \mathbf{R}^{(\mathrm{d}+1)\times 1}, \tag{7d}$$

$$\mathbf{p}_{\mathrm{g},j} = \mathrm{concat}\left( \mathbf{p}_j^1, \mathbf{p}_j^2, \ldots, \mathbf{p}_j^K \right) \in \mathbf{R}^{K(\mathrm{d}+1)\times 1}. \tag{7e}$$

where the concat operator represents the concatenation of vectors along the column axis (vertically), resulting in a column vector.

In (6), $\mathbf{x}_g$ represents the new feature vector obtained from $\mathbf{x}$ after fuzzy mapping, and $\mathbf{p}_{g,j}$ denotes the consequent parameters combined with all the rules for the $j$th output. Existing optimization technologies for linear models, e.g., the least squares method, can be used to solve for $\mathbf{p}_{g,j}$ directly. Ridge regression has been shown to be an effective optimization method [17], where the optimized consequent parameters for classification can be expressed as follows,

$$\mathbf{p}_{g,j} = \left(\lambda \mathrm{I}_{K(d+1)\times K(d+1)} + \sum_{n=1}^{N} \mathbf{x}_{gn}\left(\mathbf{x}_{gn}\right)^{\mathrm{T}}\right)^{-1} \left(\sum_{n=1}^{N} \mathbf{x}_{gn} y_{n,j}\right) \tag{8}$$

where $\mathbf{x}_{gn}$ represents the mapped input vector in the new feature space of the $n$th sample in the training set and it can be obtained by (7a)-(7c) and $\mathbf{y}_n=[y_{n,1}, \ldots, y_{n,C}]^{\mathrm{T}}$ denotes a $C$-dimensional label vector of the $n$th training sample, where $C$ represents the number of classes.

Recently, TSK FS has gained popularity for epilepsy detection. TSK FS is suitable for detecting epilepsy because of its good interpretability and strong learning ability. Yang et al. [61] proposed a TSK FS construction algorithm the basis of transductive transfer learning to solve the mismatch in distribution between the training dataset and the test dataset. Xie et al. [31] proposed a generalized hidden-mapping transductive transfer learning method, which enables transfer learning for intelligent models, including fuzzy systems. Jiang et al. [29] proposed an epileptic EEG recognition method on the basis of a multi-view learning-based TSK FS. Ni et al. [62] proposed a new noise-insensitive TSK FS based on inter-class competitive learning for EEG signal recognition. Deng et al. [19] proposed an enhanced transductive transfer learning TSK FS for epileptic EEG recognition by introducing the joint knowledge transfer mechanism. These studies demonstrate that TSK FSs have played an important role in epilepsy detection.

## 3. TSK fuzzy system based on multi-view collaborative transfer learning

A structured framework is essential for effectively implementing personalized EEG-based epilepsy detection. Typically, these frameworks consist of feature extraction, model training, and final testing modules. As the main part of the framework, this section introduces the algorithmic component, i.e., the proposed MVTL-FS. The complete detection framework is described in detail in Section 4.

### *3.1 Framework of the proposed method*

The framework of MVTL-FS is shown in Fig. 1. First, knowledge extraction is performed on the data of the source domain. Afterward, the model in the target domain is trained by the multi-view cooperative transfer learning technique. In this framework, we focus on two mechanisms: knowledge transfer from the source domain to the target domain for each view and cooperation between different views. Detailed descriptions of these mechanisms are presented below.

### *3.2 Multi-view transfer learning mechanism for fuzzy feature space*

#### *3.2.1 Multi-view distribution adaptation of fuzzy feature space*

The maximum mean discrepancy (MMD) is a common metric that is used in transfer learning to match the distributions of data between the source domain and the target domain. In this study, the MMD is used to match the distributions between the two domains in the fuzzy space for multi-view scenes.

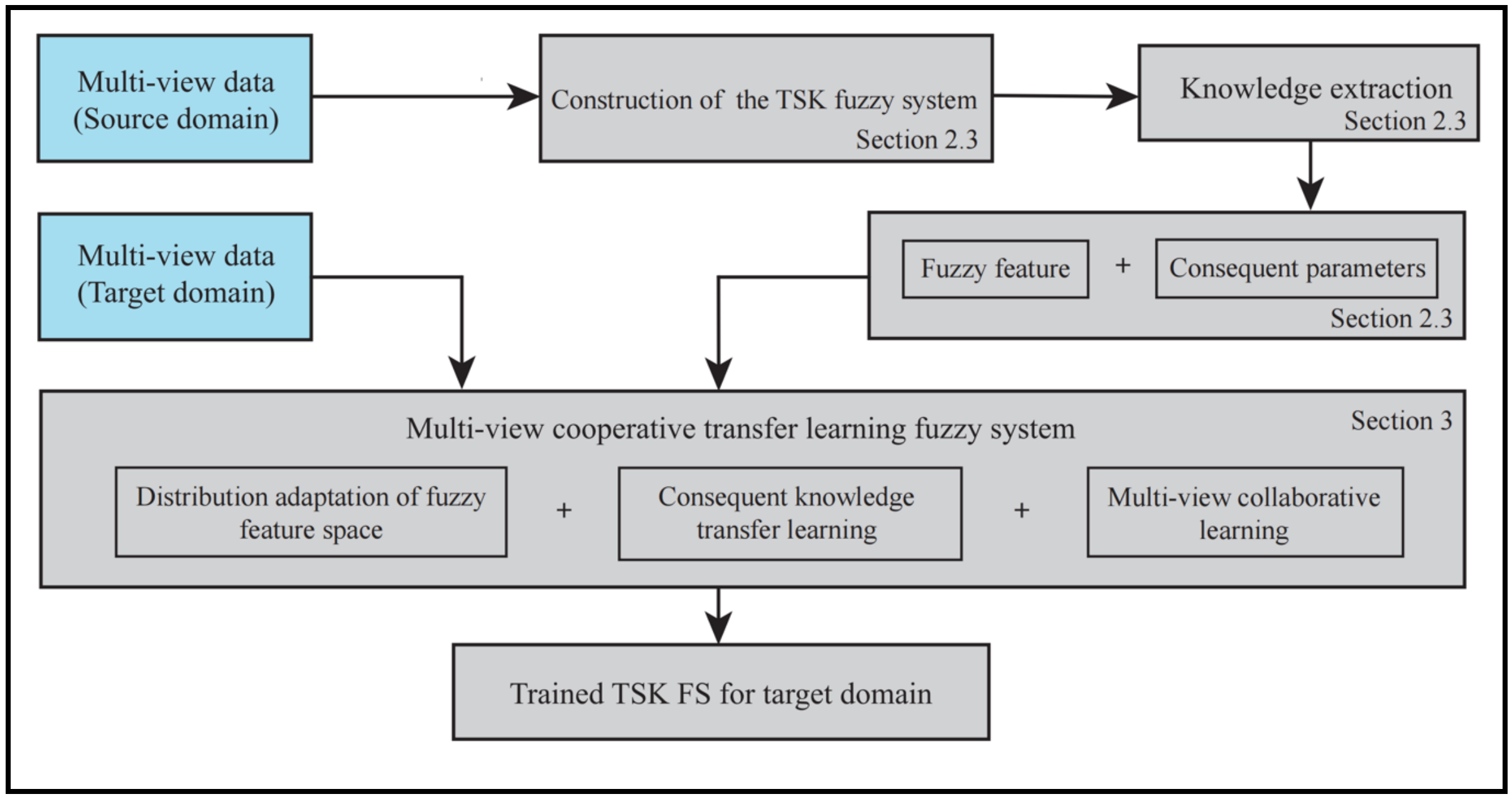

**Fig. 1.** Framework of the proposed multi-view collaborative transfer learning TSK fuzzy system.

Typically, with a source domain dataset $D_s = \{\mathbf{x}_n^v\}_{n=1}^N$ and a target domain dataset $D_t = \{\mathbf{z}_m^v\}_{m=1}^M$ under view $v$, the MMD can be expressed as follows [63]:

$$\mathrm{MMD}^{v2} = \left\| \frac{1}{N}\sum_{n=1}^{N}\phi(\mathbf{x}_n^v) - \frac{1}{M}\sum_{m=1}^{M}\phi(\mathbf{z}_m^v) \right\|^2 \tag{9}$$

where $\phi(\cdot)$ represents a mapping function used to map the original variable to the reproducing kernel Hilbert space. $N$ and $M$ denote the number of samples of the source domain and that of the target domain, respectively.

In this paper, knowledge from each view is transferred, and a mapping vector in the fuzzy space is learned based on maximum mean discrepancy (MMD) to minimize the distributional distance between the transferred view and the target. We map the source domain dataset $D_s$ and the target domain dataset $D_t$ to the fuzzy space using (7a)-(7c) to obtain $D_s' = \{\mathbf{x}_{gn}^v\}_{n=1}^N$ and $D_t' = \{\mathbf{z}_{gm}^v\}_{m=1}^M$. On the basis of (9), the MMD of two domains in the multi-view fuzzy feature space can be expressed as follows,

$$\begin{aligned} d(P_{s,map}, P_{t,map}) &= \mathrm{MMD}^2_{\boldsymbol{multi-view}} \\ &= \sum_{v=1}^{V}\sum_{j=1}^{C}\left\| \frac{1}{N}\sum_{n=1}^{N}(\mathbf{p}_{g,j}^v)^{\mathrm{T}}\mathbf{x}_{gn}^v - \frac{1}{M}\sum_{m=1}^{M}(\mathbf{p}_{g,j}^v)^{\mathrm{T}}\mathbf{z}_{gm}^v \right\|^2 \\ &= \sum_{v=1}^{V}\sum_{j=1}^{C}\Bigg( \frac{1}{N^2}\sum_{n_1=1}^{N}\sum_{n_2=1}^{N}(\mathbf{p}_{g,j}^v)^{\mathrm{T}}\mathbf{x}_{gn_1}^v(\mathbf{x}_{gn_2}^v)^{\mathrm{T}}\mathbf{p}_{g,j}^v \\ &\quad + \frac{1}{M^2}\sum_{m_1=1}^{M}\sum_{m_2=1}^{M}(\mathbf{p}_{g,j}^v)^{\mathrm{T}}\mathbf{z}_{gm_1}^v(\mathbf{z}_{gm_2}^v)^{\mathrm{T}}\mathbf{p}_{g,j}^v \\ &\quad - \frac{2}{NM}\sum_{n=1}^{N}\sum_{m=1}^{M}(\mathbf{p}_{g,j}^v)^{\mathrm{T}}\mathbf{x}_{gn}^v(\mathbf{z}_{gm}^v)^{\mathrm{T}}\mathbf{p}_{g,j}^v \Bigg) \end{aligned} \tag{10}$$

where $\mathbf{x}_{gn}^v$ represents the $n$th source domain sample of the $v$th view in the fuzzy feature space, and $\mathbf{z}_{gm}^v$ represents the $m$th target domain sample of the $v$th view in the fuzzy feature space. $\mathbf{p}_{g,j}^v$ denotes the consequent parameters of the $v$th view for the $j$th output, $V$ indicates the total number of views, and $C$ represents the total number of outputs (i.e., the number of classes in the classification dataset). Let

$$\mathbf{\Omega}_0^v = \frac{1}{N^2}\sum_{n_1=1}^{N}\sum_{n_2=1}^{N}\mathbf{x}_{gn_1}^v(\mathbf{x}_{gn_2}^v)^{\mathrm{T}} + \frac{1}{M^2}\sum_{m_1=1}^{M}\sum_{m_2=1}^{M}\mathbf{z}_{gm_1}^v(\mathbf{z}_{gm_2}^v)^{\mathrm{T}} - \frac{2}{NM}\sum_{n=1}^{N}\sum_{m=1}^{M}\mathbf{x}_{gn}^v(\mathbf{z}_{gm}^v)^{\mathrm{T}} \tag{11a}$$

$$\mathbf{\Omega}^v = \frac{\mathbf{\Omega}_0^v + \mathbf{\Omega}_0^{v\mathrm{T}}}{2} \tag{11b}$$

Equation (10) can be simplified as follows:

$$\begin{aligned} d(P_{s,map}, P_{t,map}) &= \mathrm{MMD}^2_{\boldsymbol{multi-view}} \\ &= \sum_{v=1}^{V}\sum_{j=1}^{C}(\mathbf{p}_{g,j}^v)^{\mathrm{T}}\,\mathbf{\Omega}^v\mathbf{p}_{g,j}^v \end{aligned} \tag{12}$$

*3.2.2 Comprehensive multi-view transfer mechanism*

The consequent knowledge from the source domain is further incorporated into the transfer learning process for the target domain, enabling the trained model to approximate the desired parameters more effectively under the guidance of source-domain knowledge. The knowledge transfer, which is denoted as *KT*, can be expressed as follows:

$$KT = \sum_{v=1}^{V}\sum_{j=1}^{C}(\mathbf{p}_{g,j}^v - \mathbf{p}_{g0,j}^v)^{\mathrm{T}}(\mathbf{p}_{g,j}^v - \mathbf{p}_{g0,j}^v) \tag{13}$$

where $\mathbf{p}_{g0,j}^v$ denotes the consequent parameter of the $j$th output in the source domain for view $v$. The *KT* term allows the model to implement multi-view consequent knowledge transfer to estimate the desired consequent parameters of the model in the target domain.

Based on the adaptive multi-view distribution of the fuzzy feature space and the multi-view consequent knowledge transfer, we obtain the following comprehensive multi-view transfer mechanism, which is denoted as $T$, for the construction of the TSK FS:

$$\begin{aligned} T &= \lambda_t KT + \lambda_d d(P_{s,map}, P_{t,map}) \\ &= \lambda_t\sum_{v=1}^{V}\sum_{j=1}^{C}(\mathbf{p}_{g,j}^v - \mathbf{p}_{g0,j}^v)^{\mathrm{T}}(\mathbf{p}_{g,j}^v - \mathbf{p}_{g0,j}^v) \\ &\quad + \lambda_d\sum_{v=1}^{V}\sum_{j=1}^{C}(\mathbf{p}_{g,j}^v)^{\mathrm{T}}\,\mathbf{\Omega}^v\mathbf{p}_{g,j}^v \end{aligned} \tag{14}$$

The first term in (14), which is derived from (13), transfers multi-view consequent knowledge. The second term, which is derived from (12), is to measure the distribution distance between different domains in a multi-view projected fuzzy space and to improve transfer learning performance. Ultimately, (14) can make the model transferable and improve the learning ability of the model in the target domain with multi-view data.

*3.3 Multi-view collaborative learning mechanism for tsk fuzzy systems based on fuzzy weighting and consistency constraints*

To further improve the learning ability of the constructed model, a collaborative learning mechanism involving multi-view, which is denoted as $\Psi$, is proposed and designed as follows:

$$\Psi = \frac{1}{2}\sum_{v=1}^{V}(w_v)^q\left(\sum_{j=1}^{C}\sum_{m=1}^{M}\left\|(\mathbf{p}_{g,j}^v)^T\mathbf{z}_{gm}^v - y_{mj}\right\|^2\right) + \lambda_{p_g}\sum_{v=1}^{V}\sum_{j=1}^{C}(\mathbf{p}_{g,j}^v)^{\mathrm{T}}\mathbf{p}_{g,j}^v + \frac{\lambda_{un}}{2}\sum_{v=1}^{V}\sum_{j=1}^{C}\sum_{m=1}^{M}\left\|(\mathbf{p}_{g,j}^v)^{\mathrm{T}}\mathbf{z}_{gm}^v - \frac{1}{V-1}\sum_{l=1,l\neq v}^{V}(\tilde{\mathbf{p}}_{g,j}^l)^{\mathrm{T}}\mathbf{z}_{gm}^l\right\|^2 \tag{15}$$

where $w_v$ represents the weight of the $v$th view, $q$ denotes the fuzzy index, $\mathbf{y}_m$ indicates a $C$-dimensional label vector, and $\tilde{\mathbf{p}}_{g,j}^l$ is a priori parameter vector of the fuzzy rule consequents obtained from other views except view $v$. The definitions of the other symbols are the same as those described in the previous sections.

In (15), the first two terms are derived from the traditional TSK FS for learning the consequent parameters of each independent view. The last term $(\mathbf{p}_{g,j}^v)^T\mathbf{z}_{gm}^v$ denotes the predicted $j$th output of the $v$th view, and $\frac{1}{V-1}\sum_{l=1,l\neq v}^{V}(\tilde{\mathbf{p}}_{g,j}^l)^T\mathbf{z}_{gm}^l$ denotes the average of the prior decision values of all the views, with the exception of the $v$th view. By minimizing (15), the importance of different views for the corresponding TSK FS can be obtained, and each view in the target domain can reach a consistent decision.

*3.4 Objective function based on multi-view cooperative transfer learning*

Based on (14) and (15), we propose the following objective function for training the multi-view TSK FS:

$$\min_{\mathbf{p}_{g,j}^v,\mathbf{w}} J(\mathbf{p}_{g,j}^v,\mathbf{w}) = \Psi + T \tag{16}$$

$$s.t. \sum_{v=1}^{V} w_v = 1$$

In (16), the first term is used for multi-view collaboration and the second term is used for multi-view transfer. These two terms interact to facilitate effective learning of the model parameters in the target domain.

*3.5 Optimization*

Because (16) is a nonconvex optimization problem, an alternate iteration strategy can be used for parameter optimization. This strategy has been widely applied, including in well-established methods such as FCM [64]. In recent years, fuzzy optimization strategies and neural-dynamics-based solvers have also been investigated for time-varying optimization problems, such as complex Sylvester equations [65].

**Optimize** $\mathbf{p}_{g,j}^v$ : By fixing $\mathbf{w}$, the following objective function is minimized:

$$\min_{\mathbf{p}_{g,j}^v} \frac{1}{2}\sum_{v=1}^{V}(w_v)^q\left(\sum_{j=1}^{C}\sum_{m=1}^{M}\left\|(\mathbf{p}_{g,j}^v)^T\mathbf{z}_{gm}^v - y_{mj}\right\|^2\right) + \lambda_{p_g}\sum_{v=1}^{V}\sum_{j=1}^{C}(\mathbf{p}_{g,j}^v)^{\mathrm{T}}\mathbf{p}_{g,j}^v + \frac{\lambda_{un}}{2}\sum_{v=1}^{V}\sum_{j=1}^{C}\sum_{m=1}^{M}\left\|(\mathbf{p}_{g,j}^v)^{\mathrm{T}}\mathbf{z}_{gm}^v - \frac{1}{V-1}\sum_{l=1,l\neq v}^{V}(\tilde{\mathbf{p}}_{g,j}^l)^{\mathrm{T}}\mathbf{z}_{gm}^l\right\|^2 + \lambda_t\sum_{v=1}^{V}\sum_{j=1}^{C}(\mathbf{p}_{g,j}^v - \mathbf{p}_{g0,j}^v)^{\mathrm{T}}(\mathbf{p}_{g,j}^v - \mathbf{p}_{g0,j}^v) + \lambda_d\sum_{v=1}^{V}\sum_{j=1}^{C}(\mathbf{p}_{g,j}^v)^{\mathrm{T}}\mathbf{\Omega}^v\mathbf{p}_{g,j}^v \tag{17}$$

Taking the derivative of (17) with respect to $\mathbf{p}_{g,j}^v$ and setting it to zero, the update rule for $\mathbf{p}_{g,j}^v$ is given by:

$$\mathbf{p}_{g,j}^v = \begin{pmatrix} (w_v)^q\sum_{m=1}^{M}\mathbf{z}_{gm}^v(\mathbf{z}_{gm}^v)^{\mathrm{T}} + 2\left(\lambda_{p_g} + \lambda_t\right)\mathbf{I} \\ +\lambda_{un}\sum_{m=1}^{M}\mathbf{z}_{gm}^v(\mathbf{z}_{gm}^v)^{\mathrm{T}} + 2\lambda_d(\mathbf{\Omega}^v)^{\mathrm{T}} \end{pmatrix}^{-1} \begin{pmatrix} (w_v)^q\sum_{m=1}^{M}\mathbf{z}_{gm}^v\, y_{mj} \\ +\frac{\lambda_{un}}{V-1}\sum_{l=1,l\neq v}^{V}\sum_{m=1}^{M}\mathbf{z}_{gm}^l(\mathbf{z}_{gm}^v)^{\mathrm{T}}\tilde{\mathbf{p}}_{g,j}^l + 2\lambda_t\mathbf{p}_{g0,j}^v \end{pmatrix} \tag{18}$$

**Optimize** $w_v$ : By fixing $\mathbf{p}_{g,j}^v$, the following objective function is minimized:

$$\min_{\mathbf{w}} \frac{1}{2}\sum_{v=1}^{V}(w_v)^q\left(\sum_{j=1}^{C}\sum_{m=1}^{M}\left\|(\mathbf{p}_{g,j}^v)^T\mathbf{z}_{gm}^v - y_{mj}\right\|^2\right) \tag{19}$$

$$s.t. \sum_{v=1}^{V} w_v = 1$$

Taking the derivative of (19) with respect to $w_v$ and setting it to zero, the update rule for $w_v$ is given by:

$$w_v = \frac{\left(\sum_{j=1}^{C}\sum_{m=1}^{M}\left\|(\mathbf{p}_{g,j}^v)^{\mathrm{T}}\mathbf{z}_{gm}^v - y_{mj}\right\|^2\right)^{1/1-q}}{\sum_{v'=1}^{V}\left(\sum_{j=1}^{C}\sum_{m=1}^{M}\left\|(\mathbf{p}_{g,j}^{v'})^{\mathrm{T}}\mathbf{z}_{gm}^{v'} - y_{mj}\right\|^2\right)^{1/1-q}} \tag{20}$$

The optimal $\mathbf{p}_{g,j}^v$ and $w_v$ are then obtained by iterative learning. The details are described in Algorithm 1.

*3.6 Algorithm complexity analysis*

Based on the above discussions and Algorithm 1, the computational complexity of the proposed MVTL-FS is analyzed using the Big $O$ notation. $d$ denotes the dimension of the dataset; $K$ represents the number of fuzzy rules, $S = N + M$ indicates the number of samples, where $N$ and $M$ represent the number of samples in the source domain and that in the target domain, respectively; $C$ represents the number of classes; and $T$ denotes the number of iterations. In step 2, according to (3)-(5) and (7c), the complexity of mapping the dataset into $D_s'$ and $D_t'$ is $O(2SKd)$. In step 3, the complexity of the prior consequent parameter acquisition is represented as $O(Kd(KdS+SC+d))$. Step 4 iteratively solves for the consequent parameters of different views and the weights of the views. The complexity is denoted as $O(TKd(4KdS+S^2+SC+2KdC))$ and $O(TS(KdC+SC+S^2))$ respectively. Moreover, considering the existence of different views, for $V$ views, the total complexity is $V$ times the complexity of all the steps above. Based on the analysis, Step 4 dominates the computation cost. Let

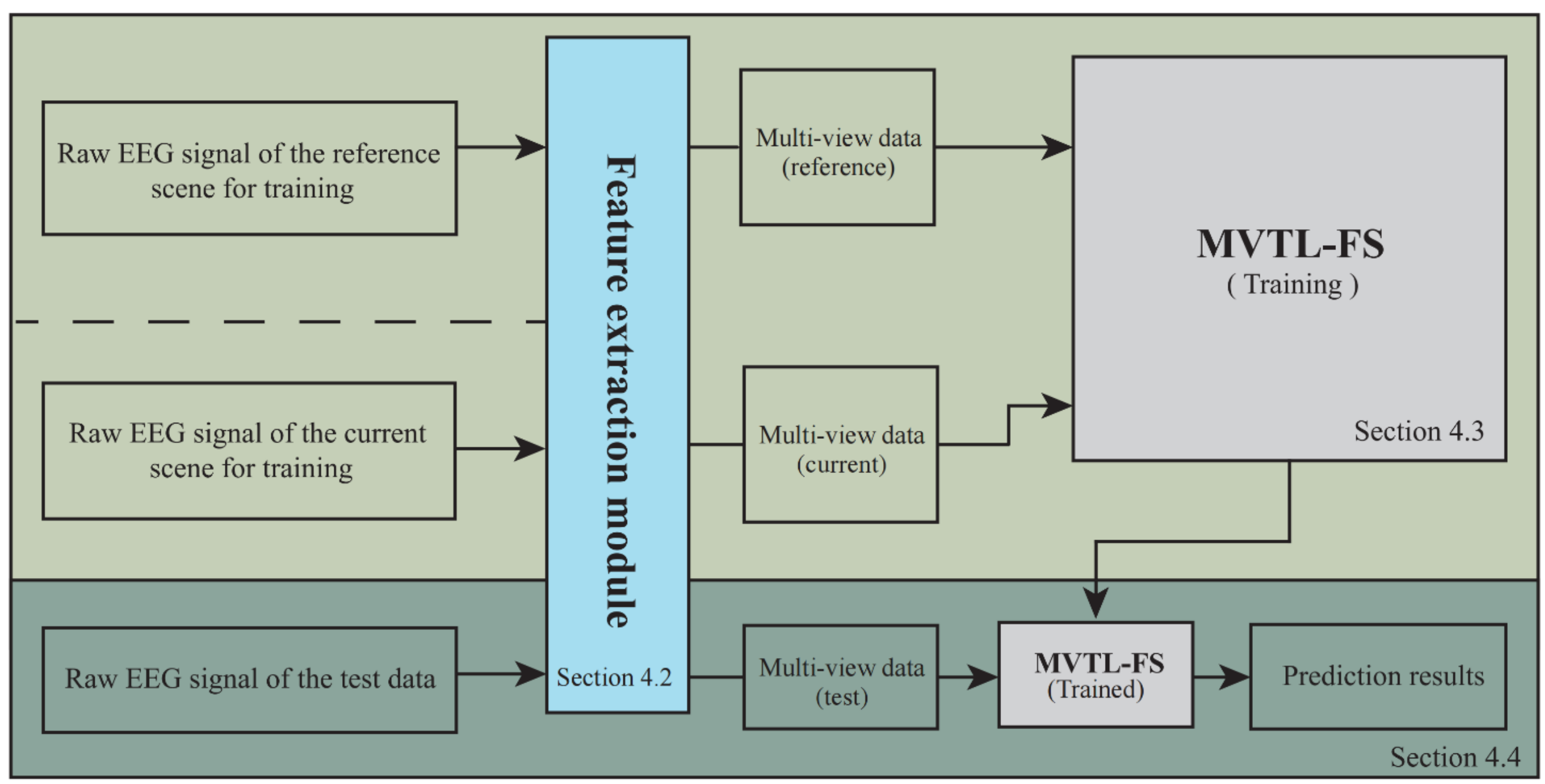

**Fig. 2.** Framework of the TSK-MVTL-FS epileptic detection system.

---

**Algorithm 1** Proposed MVTL-FS method

***Initialization*:** Set the number of fuzzy rules $K$ and the regularization parameters $\lambda_{pg}, \lambda_t, \lambda_{un}$ and $\lambda_d$. Obtain initial multi-view dataset from the source (reference) and target (current) domains. The view weight $w_v$ is $1/V$.

***Stage 1: Constructing multi-view data in the fuzzy feature space***

Step 1 The data of each view in different domains are mapped to the fuzzy inference feature space using (3)-(5) and (7a)-(7c). Deterministic clustering is used in (5) to evaluate the antecedent parameters.

Step 2 Transfer the antecedent parameters from the source domain (reference scene) to the target domain (current scene) for the construction of the corresponding fuzzy feature space. That is, we get $D_s' = \{\mathbf{x}_{gn}^v\}_{n=1}^N$ and $D_t' = \{\mathbf{z}_{gm}^v\}_{m=1}^M$.

***Stage 2: Building a TSK fuzzy system based on multi-view transfer and collaborative learning***

Step 3 Both the source domain (reference scene) parameters $\mathbf{p}_{g0}^v$ and the initial prior parameters $\tilde{\mathbf{p}}_{g,j}^l$ are obtained using (8).

Step 4 The consequent parameters of different views in the target domain (current scene) and the view weights are updated using (18), (20) and the knowledge from the source domain (reference scene).

Step 5 The final multi-view transfer model is constructed with the antecedent and consequent parameters obtained in Step 2 and Step 4.

---

a = max($Kd$, $S$) and b = max($T$, $C$, $V$), the computational cost is expressed as $O(\mathrm{a}^2\mathrm{b}^2(6\mathrm{a} + 5\mathrm{b}))$.

## 4. Epileptic EEG detection based on multi-view collaborative transfer learning TSK fuzzy system

This section provides a comprehensive overview of the TSK-MVTL-FS epilepsy detection framework, which builds on the MVTL-FS introduced earlier. Because our method is based on the TSK fuzzy system, it is named TSK-MVTL-FS to reflect this foundational aspect. TSK-MVTL-FS uses raw EEG signals as the input to extract multi-view features. These features are subsequently used to train the MVTL-FS classifier, which generates predictions for the test data. In the following subsections, we present the TSK-MVTL-FS framework, provide a detailed description of the multi-view feature extraction module, describe the training process of the MVTL-FS module, and present the testing procedure on the evaluation data.

### *4.1 Framework of epilepsy detection based on MVTL-FS*

The framework shown in Fig. 2 consists of three core components, namely, initial EEG multi-view feature extraction, training of the MVTL-FS classifier, and the test process for the future EEG signal of the test sample. The detection process is described in Algorithm 2.

---

**Algorithm 2** MVTL-FS based epilepsy EEG detection

***Stage 1: Multi-view feature extraction***

Step 1 The EEG data from existing reference scenes (source domain) and the EEG data from patients in the current scenes (target domain) are passed into the feature extraction module. Multi-view feature datasets in the time domain ($\mathbf{x}^1$, $\mathbf{z}^1$) frequency domain ($\mathbf{x}^2$, $\mathbf{z}^2$) and time-frequency domain ($\mathbf{x}^3$, $\mathbf{z}^3$) are obtained with the feature extraction module. i.e., $D_s = \{\mathbf{x}^v\}_{v=1}^3$ and $D_t = \{\mathbf{z}^v\}_{v=1}^3$.

***Stage 2: Training of the multi-view transfer learning classifier MVTL-FS***

Step 2 $D_s$ and $D_t$ are processed with the MVTL-FS module in Section 3.

Step 3 Train the MVTL-FS classifier using the algorithm in Algorithm 1.

***Stage 3: Testing***

Step 4 The multi-view dataset $T = \{\mathbf{x}_t^v\}_{v=1}^3$ is obtained with the same feature extraction module in Stage 1.

Step 5 The trained classifier in Stage 2 is tested on the dataset $T$.

---

### *4.2 Multi-view feature extraction module*

In this study, time domain features, frequency domain features, and time-frequency domain features are used as the three views for the proposed multi-view feature extraction module for epileptic EEG detection. We first obtain the EEG signal in the time domain. The corresponding feature extraction modules are described as follows:

1) Time domain features use time as a variable in the EEG signal; it can be obtained on the basis of the waveform characteristics of the original signal or the waveform characteristics of the decomposed signal. However, the spikes that appear in the signal of seizures cause signal nonlinearity, significantly increasing the linear prediction error. Typical time domain features include mean, variance and median. These parameters are often used to identify signal states [66]. In this module, the time domain features are directly inherited from the original time-varying and energy-related EEG signal, which has good temporal characteristics without extra processing.

2) Frequency domain features represent the EEG signal in terms of energy changes in frequency. In frequency domain analysis, the fast Fourier transform (FFT) [4] is typically used to convert time domain signals to the frequency domain. In this module, epileptic frequency domain features are extracted via the fast Fourier transform (FFT) from the frequency band between 4 Hz and 30 Hz [67], where features appear during seizures. Fourier transform assumes that the signal is locally smooth [68].

3) Time-frequency analysis combines the advantages of both time domain analysis and frequency domain analysis and is not limited by the assumption of locally smooth signals in the Fourier transform. A typical used method for time-frequency analysis is the wavelet transform [69], which uses a decaying orthogonal basis to model the original signal. It obtains the position of a frequency in the time domain. Among the wavelets, Daubechies (dbN) is a popular and has a high classification accuracy. In this module, we use a dbN and set the wavelet order to 4.

Finally, the multi-view feature extraction module outputs these three different types of features as three different views for subsequent training of the MVTL-FS classifier.

### *4.3 Training of MVTL-FS*

After different views of the EEG dataset are obtained, we need to train the multi-view collaborative transfer learning fuzzy system classifier, i.e., MVTL-FS. As discussed in Section 3, it is clear that MVTL-FS inherits the advantages of existing multi-view learning and transfer learning methods while overcoming the existing challenges. It makes full use of the obtained view information and reduces the impact due to insufficient data and the different distribution between the training and test data. Moreover, it has an interpretable method through fuzzy inference capability.

When the multi-view features of the data in the current scene (target domain) and the reference scene (source domain) are obtained, we first train the TSK FS for the reference scene and then extract the knowledge from the obtained models. Furthermore, the knowledge in the reference scene is used for transfer learning for the current scene. In Algorithm 1, the MVTL-FS can be trained based on the data in both scenes and the knowledge of the reference scene.

### *4.4 Testing process*

After the feature extraction module is constructed and the training of MVTL-FS is completed, they are used to classify the test samples. The overall process is described as follows: First, the test sample $\mathbf{x}_t$ is processed with the feature extraction module to obtain the required three views $\mathbf{x}_t^1$, $\mathbf{x}_t^2$ and $\mathbf{x}_t^3$. Afterwards, the test samples are fed into the trained MVTL-FS for classification and the final decision values is expressed as

$$\mathbf{f}(\mathbf{x}_{\mathrm{t}}) = [f_1(\mathbf{x}_{\mathrm{t}}), \dots, f_j(\mathbf{x}_{\mathrm{t}}), \dots, f_C(\mathbf{x}_{\mathrm{t}})] \tag{21a}$$

$$f_j(\mathbf{x}_{\mathrm{t}}) = \sum_{v=1}^{V} w_v \left(\mathbf{p}_{\mathrm{g},j}^{v}\right)^{\mathrm{T}} \mathbf{x}_{\mathrm{tg}}^{v} \tag{21b}$$

where $\mathbf{x}_{\mathrm{tg}}^{v}$ denotes the fuzzy-mapped feature vector of the test sample $\mathbf{x}_{\mathrm{t}}$ in the $v$ th view, and $f_j(\mathbf{x}_{\mathrm{t}})$ represents the decision value corresponding to the $j$ th class.

The above formulation represents a linear combination of decision values from different views. With the one-hot strategy, the final class label is expressed as $\tilde{\mathbf{y}} = [y_1, \dots, y_C]$. Let $l = \underset{1 \le j \le C}{\operatorname{argmax}} f_j(\mathbf{x}_{\mathrm{t}})$, where $l$ denotes the predicted class index. Then $y_l = 1$ and $y_j = 0$ for $j \neq l$. This indicates that $\mathbf{x}_t$ belongs to class $l$.

## 5. Experimental studies

### *5.1 Datasets*

The CHB-MIT dataset from Boston Children's Hospital is used for the experiments in this study. The dataset was collected from scalp EEG electrodes based on the International 10-20 system and was acquired at a sample rate of 256 Hz with 16-bit resolution. The dataset is available on the Physionet website (https://archive.physionet.org/pn6/chbmit/).

To evaluate the performance of MVTL-FS, we use the EEG signals of the first five subjects for performance comparison, denoted as S1, S2, S3, S4, and S5, respectively. The corresponding datasets to be generated are D1, D2, D3, D4, and D5. For each subject, based on the EEG signal at seizure and without a seizure, positive samples and negative samples are extracted respectively. In our experiments, each sample corresponds to one second of the EEG signal segment, i.e., 256 sampling points for each channel. To prevent overfitting, a portion of the normal state signal (without seizure) is discarded in the experiment, and a sliding window technique is used to increase the number of epileptic seizure samples. The sliding window uses a fixed length of the EEG signal sampled within one second and oversampling is implemented by overlapping windows. Finally, the five datasets D1-D5 are generated for the experiments.

**Table 1**
Datasets used in the experiments.

| ID | Total number of samples | Number of seizure samples | Number of non-seizure samples |
|---|---|---|---|
| P1 | 268 | 108 | 160 |
| P2 | 103 | 37 | 66 |
| P3 | 258 | 89 | 169 |
| P4 | 224 | 73 | 151 |
| P5 | 336 | 108 | 228 |

**Table 2**
Performance of MVTL-FS on different datasets (%).

| Datasets (source) | P2 | P3 | P4 | P5 | P1 | P3 | P4 | P5 | P1 | P2 | P4 | P5 |
|---|---|---|---|---|---|---|---|---|---|---|---|---|
| Datasets (target) | P1 | | | | P2 | | | | P3 | | | |
| Accuracy | 98.28 | 98.62 | 98.58 | 98.45 | 98.33 | 97.74 | 97.76 | 97.89 | 98.16 | 98.08 | 98.01 | 98.24 |
| SD | 0.0126 | 0.0097 | 0.0099 | 0.0089 | 0.0107 | 0.0234 | 0.0121 | 0.0144 | 0.0103 | 0.0125 | 0.0089 | 0.0090 |
| Average | **98.48 (0.0102)** | | | | **97.93 (0.0158)** | | | | **98.12 (0.0101)** | | | |
| Datasets (source) | P1 | P2 | P3 | P5 | P1 | P2 | P3 | P4 | | | | |
| Datasets (target) | P4 | | | | P5 | | | | | | | |
| Accuracy | 96.02 | 95.59 | 95.91 | 95.87 | 96.28 | 95.80 | 96.55 | 96.26 | | | | |
| SD | 0.0205 | 0.0218 | 0.0145 | 0.0161 | 0.0097 | 0.0129 | 0.0072 | 0.0136 | | | | |
| Average | **95.85 (0.0182)** | | | | **96.22 (0.0113)** | | | | | | | |

SD denotes standard deviation and is also inside the brackets.

**Table 3a**
Performance comparison of no multi-view methods (%).

| Dataset | Time Domain | | Frequency Domain | | Time-Frequency Domain | | EEGnet [74] | MVTL-FS+ |
|---|---|---|---|---|---|---|---|---|
| | TSK FS [53] | Tradaboost+ [73] | TSK FS [53] | Tradaboost+ [73] | TSK FS [53] | Tradaboost+ [73] | | |
| P1 | 76.01<br>(0.0268) | 61.87<br>(0.0923) | 98.21<br>(0.0104) | 95.29<br>(0.0288) | 79.32<br>(0.0266) | 73.26<br>(0.0392) | 94.84<br>(0.0117) | **98.48<br>(0.0102)** |
| P2 | 76.06<br>(0.0551) | 64.05<br>(0.0652) | 95.93<br>(0.0246) | 88.49<br>(0.1260) | 77.39<br>(0.0383) | 70.37<br>(0.0276) | 87.11<br>(0.0370) | **97.93<br>(0.0158)** |
| P3 | 74.51<br>(0.0284) | 63.65<br>(0.0797) | 95.92<br>(0.0282) | 91.40<br>(0.0466) | 85.63<br>(0.0260) | 78.22<br>(0.0530) | 93.97<br>(0.0144) | **98.12<br>(0.0101)** |
| P4 | 70.22<br>(0.0244) | 58.76<br>(0.0733) | 89.89<br>(0.0236) | 78.78<br>(0.0458) | 79.04<br>(0.0234) | 66.98<br>(0.0447) | 77.59<br>(0.0153) | **95.85<br>(0.0182)** |
| P5 | 73.26<br>(0.0323) | 59.87<br>(0.0763) | 94.72<br>(0.0155) | 87.22<br>(0.0541) | 82.53<br>(0.0161) | 75.08<br>(0.0296) | 90.25<br>(0.0140) | **96.22<br>(0.0113)** |
| Average | 74.01<br>(0.0242) | 61.64<br>(0.0231) | 94.93<br>(0.0309) | 88.24<br>(0.0613) | 80.78<br>(0.0329) | 72.78<br>(0.0432) | 88.75<br>(0.0696) | **97.32<br>(0.0120)** |

+ For transfer learning-based methods, the results are the average of the accuracies based on different source domains.
The standard deviation is inside the bracket.

**Table 3b**
Performance comparison of multi-view methods (%).

| Dataset | MV-TSK-FS [29] | AMVMED [70] | mulEEG+[71] | MMNET [72] | MVTL-FS* ($\lambda_t = 0$, $\lambda_d = 0$) | MVTL-FS+ |
|---|---|---|---|---|---|---|
| P1 | 96.60<br>(0.0155) | 68.26<br>(0.0272) | 93.18<br>(0.0081) | 91.94<br>(0.0614) | 96.15<br>(0.0210) | **98.48<br>(0.0102)** |
| P2 | 94.00<br>(0.0293) | 69.88<br>(0.0426) | 97.41<br>(0.0076) | 94.31<br>(0.0279) | 94.18<br>(0.0375) | **97.93<br>(0.0158)** |
| P3 | 96.32<br>(0.0166) | 69.31<br>(0.0259) | 85.68<br>(0.0360) | 89.19<br>(0.0059) | 96.17<br>(0.0151) | **98.12<br>(0.0101)** |
| P4 | 92.09<br>(0.0298) | 66.34<br>(0.0201) | 79.20<br>(0.0246) | 91.78<br>(0.0283) | 92.60<br>(0.0191) | **95.85<br>(0.0182)** |
| P5 | 93.64<br>(0.0192) | 67.40<br>(0.0258) | 93.41<br>(0.0125) | 94.94<br>(0.00003) | 92.72<br>(0.0209) | **96.22<br>(0.0113)** |
| Average | 94.53<br>(0.0190) | 68.24<br>(0.0143) | 89.78<br>(0.0727) | 92.43<br>(0.0229) | 94.36<br>(0.0175) | **97.32<br>(0.0120)** |

* MVTL-FS ($\lambda_t = \lambda_d = 0$) is a version of the proposed MVTL-FS in which the transfer learning abilities are disabled.
+ For transfer learning-based methods, the results are the average of the accuracies based on different source domains.
The standard deviation is inside the bracket.

*5.2 Experimental settings*

To simulate the scene of personalized epileptic detection when the training data are insufficient, 5% of the samples in datasets D1, D2, …, D5 are used in our experiments. These samples are denoted as P1, P2, …, P5. The total number of samples, the number of seizure samples, and the number of non-seizure samples are listed in Table 1. To implement transfer learning, 20 different transfer tasks are constructed by selecting any two from the five datasets as the reference scene (source) or the current scene (target), e.g., P1→P2, P1→P3, …, P5→P4. All the datasets in the experiments are normalized.

Five multi-view learning methods and three non-multi-view learning methods are used for comparison. The five multi-view learning algorithms used are MV-TSK-FS [29], AMVMED [70], mulEEG [71], MMNET [72] and MVTL-FS ($\lambda_t$=0, $\lambda_d$=0), where the last algorithm MVTL-FS ($\lambda_t$=0, $\lambda_d$=0) denotes the proposed MVTL-FS without transfer learning ability. In addition, we have included two multi-view deep learning methods (mulEEG and MMNET) for a more comprehensive comparison, where mulEEG has the capability of transfer learning. To remain consistent with the original design of these models, both methods used only the time domain and frequency domain views. The non-multi-view learning algorithms adopted include the traditional TSK fuzzy system (TSK FS) [53], Transfer AdaBoost based on the TSK FS learner (TrAdaBoost (TSK)) [73], as well as EEGNet [74], a widely used deep learning baseline for EEG-based seizure detection. Among the above methods, all the traditional machine learning algorithms are related to fuzzy systems, except AMVMED.

To validate the proposed MVTL-FS, five-fold cross-validation is used. The number of iterations is set to 10 for all the algorithms except AMVMED which do not require iterations. The trade-off parameters and regularization parameters of the algorithms are optimally set by searching the grid {0.01, 0.1, 1, 10, 100}. With respect to the algorithms that use the TSK fuzzy system, the number of rules is 3. The fuzzy index of the weighting term of the proposed MVTL-FS is optimized by searching the grid {0.25, 0.5, 1, 2, 4}. For deep learning–based methods, the experimental settings and hyperparameters are set in accordance with the original publications.

The performance index in the experiments is accuracy, which indicates the percentage of correctly classified test samples.

### *5.3 Results analysis*

Table 2 shows the accuracy of the proposed MVTL-FS on different tasks for various current scenes (target domains) and reference scenes (source domains). Tables 3a and 3b show the accuracy of all the algorithms on the CHB-MIT datasets. In the experiments, non-multi-view algorithms are applied for each of the different views (time domain, frequency domain, and time-frequency domain) respectively. With respect to the transfer learning-based methods, i.e., Tradaboost, mulEEG, and MVTL-FS, the average of the accuracies based on different source domains are provided. The following observations can be made:

1) Table 3a shows that the same single view algorithm yields different results under different views. In general, the best results are obtained in the frequency domain, whereas the worst results are obtained in the time domain. Although the TSK FS achieves competitive performance in the frequency domain, this advantage is confined to a single-view setting and does not generalize across different views. In addition, the Tradaboost algorithm, which is based on the TSK FS, exhibits a certain degree of negative transfer when compared to using the TSK FS alone. EEGNet, included as a deep learning baseline under the time domain view, improves performance compared with traditional fuzzy-system-based methods due to its stronger nonlinear feature extraction capability. However, its performance remains inferior to that of the proposed MVTL-FS across all datasets, indicating that even models with powerful nonlinear representation capacity are constrained when relying on a single view, whereas MVTL-FS benefits from exploiting complementary information from multiple views together with transfer learning.

Overall, the results in Table 3a indicate that the limited performance of non-multi-view methods mainly stems from the restricted information provided by a single feature view, which is insufficient to capture the complex temporal–spectral characteristics of epileptic EEG signals and thus limits further performance improvement.

2) The algorithms that are related to fuzzy systems, i.e., TSK FS, Tradaboost (TSK), MV-TSK FS, and MVTL-FS ($\lambda_t$=0, $\lambda_d$=0), are considered. The multi-view algorithms MV-TSK-FS and MVTL-FS ($\lambda_t$=0, $\lambda_d$=0) have good performance compared to non-multi-view algorithms. Therefore, considering the available views can effectively avoid the instability caused by considering only a specific view. Moreover, the performance can be improved by the collaboration of the views.

3) In Table 3b, we compare all the multi-view learning methods. The experimental results indicate that our proposed method, MVTL-FS, achieved view enhancement by incorporating a transfer learning mechanism, outperforming the other algorithms. Moreover, when the transfer learning ability was removed (i.e., MVTL-FS ( $\lambda_t$=0, $\lambda_d$=0 )), the model performance decreased. These findings indicate that the transfer learning mechanism is effective at improving the performance of our method. These results further reveal that although existing multi-view baseline methods can integrate different features, their generalizability performance remains limited because of the lack of an effective cross-domain adaptation mechanism. Furthermore, although deep learning methods typically perform well, they often require a large amount of data to achieve optimal results. Additionally, their strong performance is often built on structural designs specific to the target task. As a result, these methods did not achieve the best performance in this study.

Combining the results from Table 3a and Table 3b, we conclude that the comprehensive integration of multi-view information, together with the incorporation of a transfer learning mechanism, plays a significant role in enhancing model performance. These findings highlight the effectiveness of combining multi-view learning with transfer learning in EEG-based epilepsy detection.

In summary, the proposed MVTL-FS achieves superior performance. The experimental results indicate that MVTL-FS reduces performance fluctuations caused by different views. The proposed MVTL-FS has a distinct advantage over the other algorithms because of the knowledge applied from the source domain and the collaboration of different views simultaneously.

Although the above results demonstrate the effectiveness of the proposed MVTL-FS, EEG-based epileptic seizure detection remains an active and rapidly evolving research area [75, 76]. To provide a broader comparison, we additionally provide a reference-only comparison summarizing representative deep

**Table 4**

Reference-only comparison with representative recent deep learning methods for EEG seizure detection on the CHB-MIT dataset.

| Method | Year | Details of the method | Metric (Acc %) |
|---|---|---|---|
| Lian et al. [77] | 2023 | Graph Transformer Network (graph-based inter-channel modeling) | 98.43 |
| Zhao et al. [78] | 2023 | Hybrid Attention Network (GAT + Transformer for spatial–temporal modeling) | 98.30 |
| Wang et al. [79] | 2024 | Unsupervised Domain Adaptation Network (CNN + adversarial domain alignment) | 91.22 |
| Lee et al. [80] | 2024 | ResNet–LSTM Hybrid Network (supervised contrastive representation learning) | 91.90 |
| Tripathy et al. [81] | 2025 | Hybrid SLT–SNN Framework (slantlet transform + spiking neural network) | 93.00 |

**Table 5**

Performance of MVTL-FS on all datasets (%).

| Datasets (source) | P13 | P9 | P4 | P22 | P22 | P3 | P22 | P5 | P2 | P19 | P8 | P1 |
|---|---|---|---|---|---|---|---|---|---|---|---|---|
| Datasets (target) | P1 | P2 | P3 | P4 | P5 | P6 | P7 | P8 | P9 | P10 | P11 | P13 |
| Accuracy | 99.62 | 96.10 | 98.01 | 99.11 | 96.13 | 96.84 | 96.41 | 95.35 | 97.22 | 96.63 | 97.12 | 95.16 |
| SD | 0.0084 | 0.0416 | 0.0089 | 0.0122 | 0.0081 | 0.0471 | 0.0140 | 0.0132 | 0.0278 | 0.0363 | 0.0135 | 0.0464 |

| Datasets (source) | P22 | P1 | P22 | P22 | P24 | P14 | P11 | P19 | P14 | P22 | P2 | Average |
|---|---|---|---|---|---|---|---|---|---|---|---|---|
| Datasets (target) | P14 | P15 | P16 | P17 | P18 | P19 | P20 | P21 | P22 | P23 | P24 | |
| Accuracy | 97.05 | 93.31 | 97.50 | 94.58 | 96.91 | 95.24 | 95.68 | 95.00 | 97.69 | 98.43 | 95.11 | 96.53 |
| SD | 0.0445 | 0.0222 | 0.0559 | 0.0381 | 0.0281 | 0.0576 | 0.0363 | 0.0456 | 0.0211 | 0.0164 | 0.0255 | 0.0151 |

**Table 6**

Comparison of algorithm performance with different sizes of datasets (%).

| Size of datasets | MV-TSK-FS [29] | AMVMED [70] | mulEEG [71] | MMNET [72] | MVTL-FS ($\lambda_t = \lambda_d = 0$) | TSK (all view) [53] | Tradaboost (all view) [73] | EEGnet [74] | MVTL-FS |
|---|---|---|---|---|---|---|---|---|---|
| 5% | 94.53 (0.0190) | 68.24 (0.0143) | 89.78 (0.0727) | 92.43 (0.0229) | 94.36 (0.0175) | 83.24 (0.1067) | 74.22 (0.1336) | 88.75 (0.0696) | **97.32 (0.0120)** |
| 10% | 95.51 (0.0291) | 67.88 (0.0104) | 89.03 (0.0762) | 94.39 (0.0313) | 95.78 (0.0315) | 83.56 (0.1146) | 73.26 (0.1467) | 90.42 (0.0585) | **97.03 (0.0221)** |
| 20% | 95.23 (0.0247) | 67.95 (0.0159) | 86.29 (0.0792) | 97.57 (0.0347) | 96.49 (0.0176) | 84.26 (0.1121) | 72.82 (0.1638) | 90.69 (0.0562) | **97.72 (0.0137)** |
| 30% | 96.85 (0.0207) | 63.31 (0.0643) | 86.12 (0.0775) | 97.11 (0.0260) | 96.86 (0.0217) | 84.67 (0.1103) | 72.04 (0.1806) | 91.13 (0.0562) | **98.10 (0.0106)** |
| Average | 95.53 (0.0097) | 66.85 (0.0236) | 87.60 (0.0161) | 96.36 (0.0172) | 95.87 (0.0110) | 83.93 (0.0065) | 73.09 (0.0091) | 90.25 (0.0104) | **97.54 (0.0047)** |

learning–based methods reported in the past three years on the CHB-MIT dataset. Although all these studies are based on the same dataset, differences in research objectives and experimental settings lead to different reported results. Therefore, the reported results are provided for reference only rather than direct quantitative comparison (see Table 4).

Furthermore, to validate the robustness of the proposed method, we extend the target domain to all patients in the CHB-MIT dataset. The patient with the closest Euclidean distance to the target patient is selected as the source in the experiment. The rest of the settings are consistent with the above experiments. Patient P12 has been excluded as the target domain for experiments because the montage was changed during the detection of the EEG [64]. The experimental results are presented in Table 5, which show that the proposed method still achieves highly competitive average performance, consistent with the results reported in Tables 3a and 3b.

### *5.4 Model analysis*

#### *5.4.1 Effect of sample size*

In the previous experiments, we constructed datasets P1-P5 by extracting 5% of the samples from the original datasets D1-D5, respectively. To determine the effect of sample size on the stability of the proposed method, we vary the sample size of the datasets P1-P5 by extracting 10%, 20% and 30% of the samples for the original datasets D1-D5.

The experimental results are provided in Table 6. For the single-view algorithm, we take the average of the results of all the views. It can be seen that the performance of the algorithms fluctuates with the size of the training datasets. Among the algorithms of the same mechanism, the multi-view methods outperform the non-multi-view methods. In addition, when the sample size changes, the performance of the MVTL-FS always outperforms that of the other algorithms. MVTL-FS achieves the best result when the performance of the algorithms on all the tasks on average is considered.

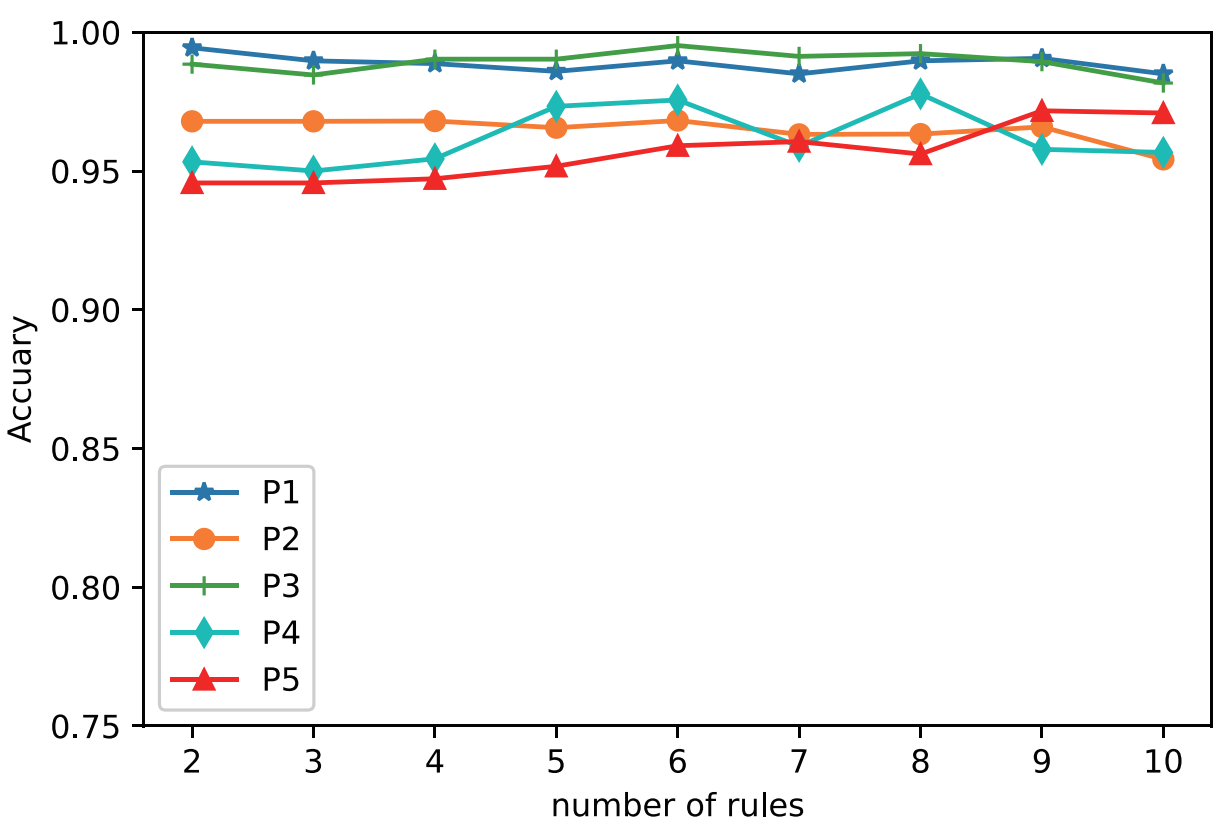

**Fig. 3.** Effect of the number of rules

*5.4.2 Effect of the number of fuzzy rules*

The number of fuzzy rules in the TSK FS affects the performance of the proposed MVTL-FS. It is desirable to use as few rules as possible to improve the performance of the algorithm [82]. We analyze the effect by varying the number of fuzzy rules from 2 to 10 and run the experiments on the constructed datasets P1-P5. The results in Fig. 3 show that the performance of MVTL-FS is relatively stable for different numbers of rules. The average accuracy is the highest when the number of rules is 6, but the advantage is not significantly better than the other settings.

The results demonstrate that the proposed MVTL-FS method achieves high accuracy with only a small number of fuzzy rules, thereby enabling the construction of a compact and interpretable model while maintaining strong classification performance. In our experiments, the number of rules is set to 3. Furthermore, based on the complexity analysis in Section 3.6, the computational complexity of MVTL-FS increases greatly with the number of rules. Hence, it is necessary to use fewer rules.

*5.4.3 Parameter sensitivity and convergence analysis*

To analyze the sensitivity of the proposed method on the parameters $\lambda_{p_g}$, $\lambda_t$, $\lambda_d$ and $\lambda_{un}$, we evaluate the classification performance by varying the parameters within the range $\{10^{-4}, 10^{-3}, 10^{-2}, 10^{-1}, 10^{0}, 10^{1}, 10^{2}, 10^{3}, 10^{4}\}$. Taking the dataset P2 as an example, Figs. 4(a) to 4(d) show the classification accuracy under different parameter settings in different multi-view transfer tasks. Fig. 4(a) shows that the classification accuracy changes significantly when the value of $\lambda_{p_g}$ is greater than 1. Fig. 4(b) shows that stable classification performance can be obtained when $\lambda_t$ is less than 1. It can be seen from Fig. 4(c) that the proposed method is not sensitive to

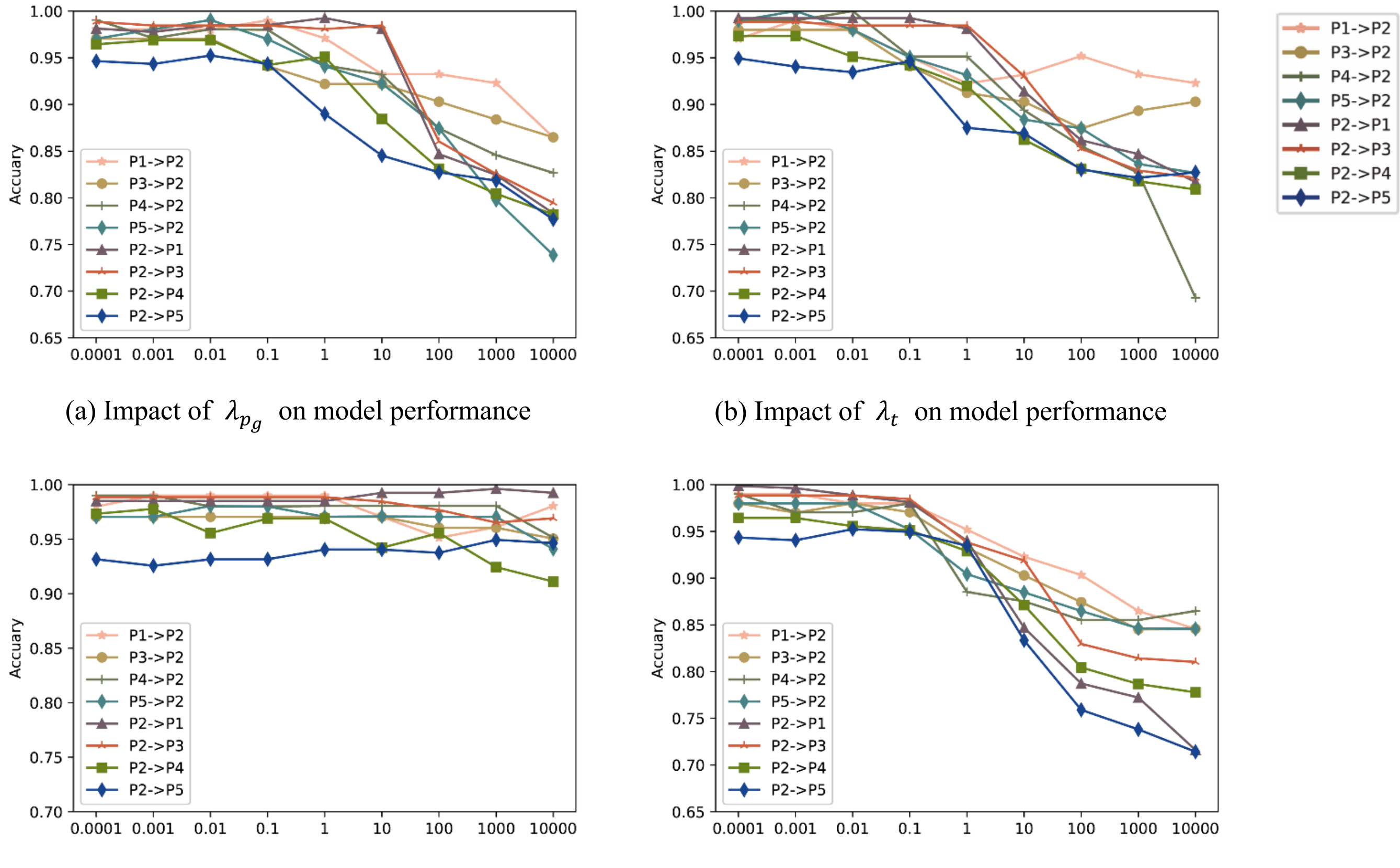

(a) Impact of $\lambda_{p_g}$ on model performance

(b) Impact of $\lambda_t$ on model performance

(c) Impact of $\lambda_d$ on model performance

(d) Impact of $\lambda_{un}$ on model performance

**Fig. 4.** Parameter sensitivity analysis.

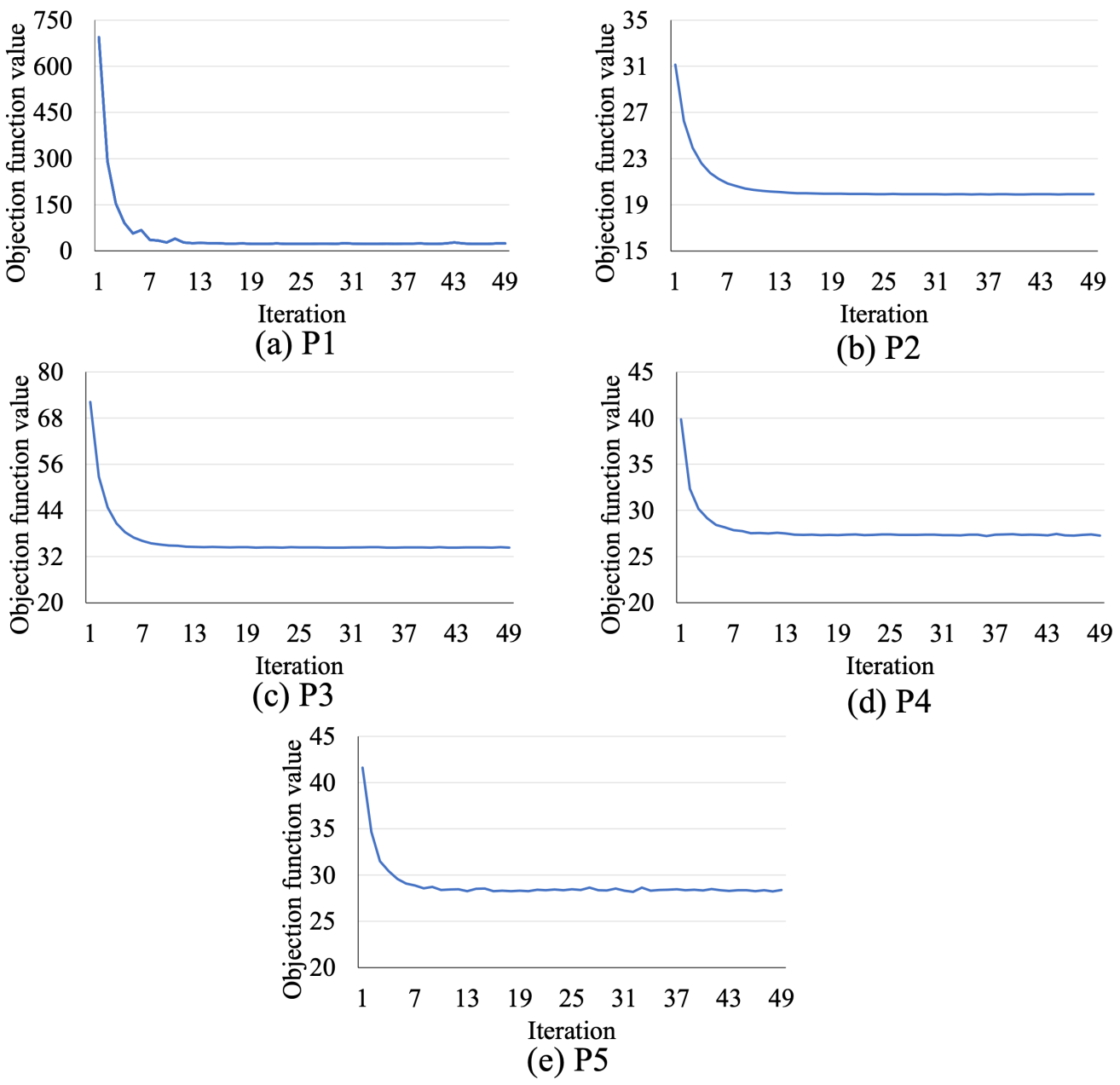

**Fig. 5.** Convergence curves of the proposed method on different patient datasets.

the change of $\lambda_d$. Meanwhile, from Fig. 4(d), we can see that the proposed method is sensitive to $\lambda_{un}$.

The convergence curves of the proposed MVTL-FS for each dataset are shown in Fig. 5. The proposed method achieves good convergence. Over time, the value of the objective function decreases rapidly and eventually stabilizes.

**Table 7**
Classification performance for different view combinations (%)

| | V2+V3 | V1+V2 | V1+V3 | V1+V2+V3 |
|---|---|---|---|---|
| P1 | 97.93 | 97.98 | 84.37 | **98.48** |
| P2 | 97.39 | **98.56** | 88.17 | 97.93 |
| P3 | 97.66 | 97.73 | 89.71 | **98.12** |
| P4 | **96.69** | 94.78 | 84.45 | 95.85 |
| P5 | 95.92 | 96.02 | 86.04 | **96.22** |

### *5.4.4 Multi-view analysis*

In this subsection, we analyze the effects of different view combinations on model performance. In the experiments, the views were defined as V1 (time domain), V2 (frequency domain), and V3 (time-frequency domain), and they were combined into four configurations: V2+V3, V1+V3, V1+V2, and V1+V2+V3. In the experiments, the model framework and parameter settings were consistent with those of the main experiments. The experimental results are presented in Table 7.

Table 7 presents the classification accuracies for different view combinations. Overall, compared with the single-view methods in Table 3a, the multi-view methods consistently achieved excellent performance on the same patients. Further analysis reveals performance variations across different dual-view combinations: V1+V3 shows relatively poor performance, whereas the difference between V1+V2 and V2+V3 remains comparatively small. Overall, the combination of the three views provides more stable and superior results in most tasks, indicating complementarity among the different views. Hence, to avoid the additional complexity introduced by view selection, adopting a comprehensive multi-view combination, when feasible, represents a more effective strategy for enhancing the model's generalization capability.

**Table 8**
Rule base of the overall model.

| | |
|---|---|
| **Rule 1:** | |
| **IF:** | the 1*st* feature is *High,* and |
| | the 2*nd* feature is *High,* and |
| | the *3rd* feature is *Middle,* and |
| | … |
| **Then:** | the 1*st* output is 0.0299+0.0166$x_1$+0.0020$x_2$+…+0.0278$x_d$, and |
| | the 2*nd* output is 0.0029-0.0045$x_1$+0.0016$x_2$+…+0.0043$x_d$ |
| **Rule 2:** | |
| **IF:** | the 1*st* feature is *Low,* and |
| | the 2*nd* feature is *Low,* and |
| | the *3rd* feature is *Low,* and |
| | … |
| **Then:** | the 1*st* output is -1.410-0.720$x_1$+0.174$x_2$+…+1.495$x_d$, and |
| | the 2*nd* output is 1.424+0.775$x_1$-0.163$x_2$+…-0.902$x_d$ |
| **Rule 3:** | |
| **IF:** | the 1*st* feature is *Middle*, and |
| | the 2*nd* feature is *Middle*, and |
| | the *3rd* feature is *High*, and |
| | … |
| **Then:** | the 1*st* output is 0.0234+0.0166$x_1$+0.0008$x_2$+…+0.0274$x_d$, and |
| | the 2*nd* output is 0.0043+0.0030$x_1$+0.0002$x_2$+…+0.0050$x_d$ |

### *5.4.5 Interpretability analysis*

The transparency of the fuzzy system (FS) is attributed primarily to the rule-based fuzzy inference mechanism. To demonstrate the transparency of the proposed MVTL-FS, we provide an example of how to generate IF-THEN rules for the patient frequency view.

In the TSK fuzzy system, each fuzzy set $A_i^k$ for rule $k$ and dimension $i$ can be interpreted in linguistic terms. In our experiments, because the rule number is 3, their central language can be described as low, middle, or high. The provided linguistic description is merely one possible example of the IF part of a fuzzy rule, because different medical experts may interpret fuzzy rules differently. In our example, the rule base of the whole model is shown in Table 8. Because Gaussian membership functions are employed, the corresponding fuzzy sets are shown in Fig. 6.

Taking the first row as an example, the center value and variance of the first dimension in the first fuzzy rule are 0.9637 and 0.0261, respectively. The center value "0.9637" ranks first among the three center values (i.e., 0.9637, 0.2147, and 0.7928). Consequently, this fuzzy set is assigned the linguistic term "High". After all fuzzy sets are assigned linguistic terms, the fuzzy system can be interpreted in terms of fuzzy rules.

Specifically, Rule 1 can be described in linguistic terms as follows:

**IF** the 1st feature (i.e., the energy of the EEG signal in the 1st dimension) is ***High***, and

the 2nd feature (i.e., the energy of the EEG signal in the 2nd dimension) is ***High***, and

the 3rd feature (i.e., the energy of the EEG signal in the 3rd dimension) is ***Middle***, and

…;

**THEN** this rule gives the decision values of two outputs with the following formula:

$$f^1(\mathbf{x}) = [f_1^1(\mathbf{x}), f_2^1(\mathbf{x})] = \begin{bmatrix} 0.0299 + 0.0166x_1 + 0.0020x_2 + \cdots + 0.0278x_d, \\ 0.0029 - 0.0045x_1 + 0.0016x_2 + \cdots + 0.0043x_d \end{bmatrix}, \quad (22)$$

where, $f_1^1(\mathbf{x})$ represents the output of the first class, and $f_2^1(\mathbf{x})$ represents the output of the second class.

Similarly, the linguistic expressions for other rules and their corresponding outputs, $f^2(\mathbf{x})$ and $f^3(\mathbf{x})$ can also be derived. Ultimately, the global output for the current view is obtained as $f(\mathbf{x}) = f^1(\mathbf{x}) + f^2(\mathbf{x}) + f^3(\mathbf{x})$.

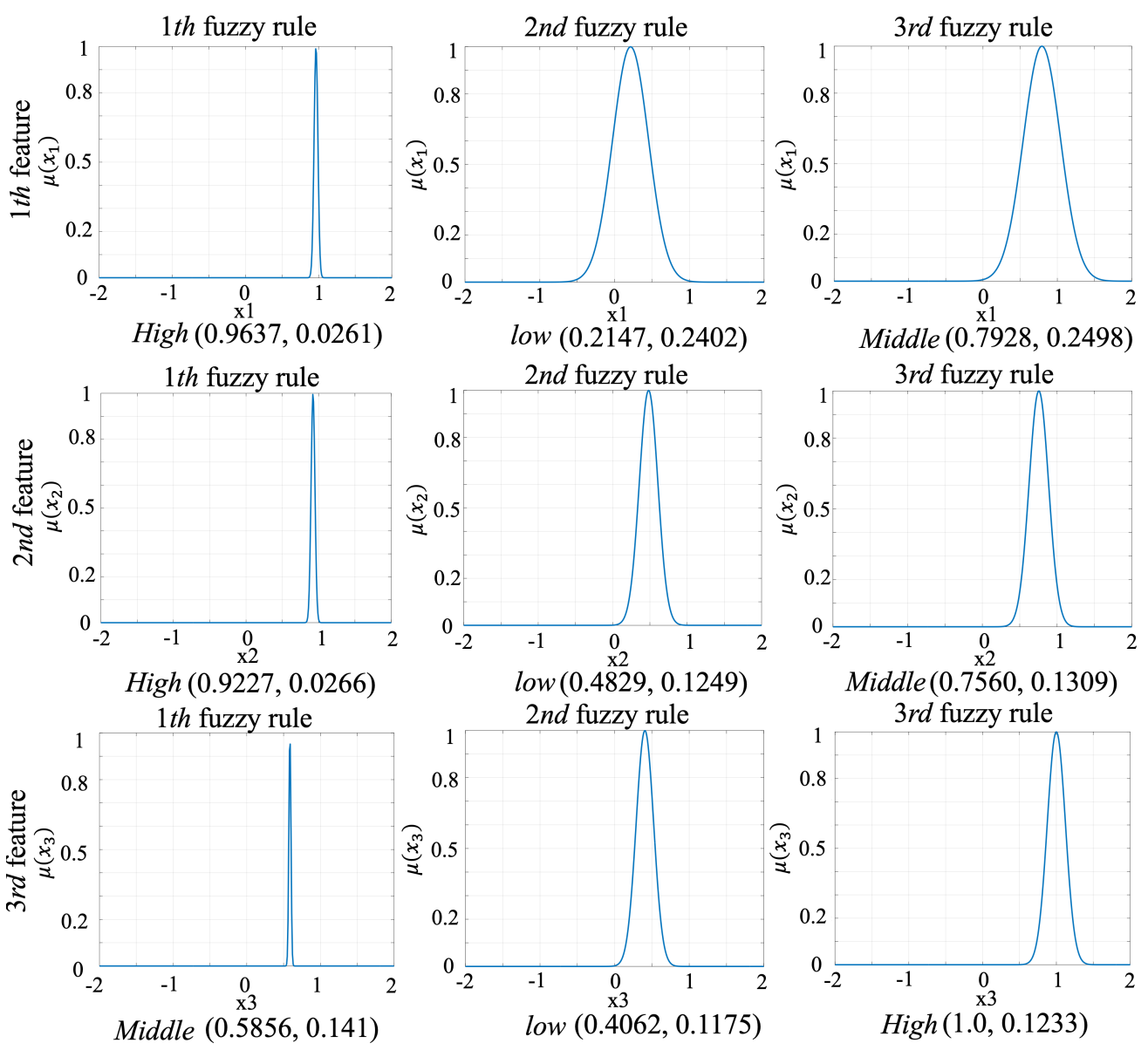

**Fig. 6.** The membership functions and the possible linguistic explanation of each fuzzy set for the first, second, and third dimension in the fuzzy system associated with frequency domain view that is trained by the proposed MVTL-FS method.

## 6. Conclusion

### *6.1 Concluding remarks*

In this paper, we propose a TSK fuzzy system that incorporates multi-view collaborative transfer learning abilities for epilepsy EEG detection. It integrates multi-view learning, transfer learning, and fuzzy system to enhance the detection performance. The proposed MVTL-FS exploits the information from different views and applies transfer learning for each view to address the issues caused by the insufficient data for personalized diagnosis. Experiments have shown that compared with the other methods, MVTL-FS is a stable method with higher classification accuracy. These results reveal not only the performance advantages of the proposed method but also its potential theoretical and clinical value. From a neuroscience perspective, the proposed multi-view features can identify epileptic seizure–related physiological patterns, such as abnormal synchronous discharges, from multiple dimensions, whereas the TSK fuzzy rules provide interpretability. These findings provide more empirical support for elucidating the electrophysiological mechanisms of seizures and inter-individual variability. Clinically, the accuracy of the method remains stable even with a limited number of individual samples, making it suitable for automated EEG-based seizure warning and rapid triage, reducing the burden of manual review and the risk of missed diagnoses.

### *6.2 Limitation of study*

Although the proposed method performs well in epilepsy EEG detection, several limitations should be acknowledged. First, although the use of fuzzy rules increase the representational capacity of the model, they inevitably increase the computational cost, which may limit the scalability of large-scale data processing or real-time applications. Second, the current transfer learning framework is limited to within-view adaptation and has not yet been extended to cross-view scenarios, which limits its applicability in more complex and heterogeneous clinical settings. Third, the current framework, based on a type 1 TSK fuzzy system, has proven effective in the present task by providing interpretability and the ability to manage uncertainty to a certain extent, but may have limited capacity to address more complex clinical uncertainties. These limitations highlight important directions for future work.

### *6.3 Future work*

Future research will proceed in several directions. First, we will investigate more efficient rule-reduction and lightweight inference mechanisms to reduce computational overhead and improve applicability to large-scale or real-time tasks. Second, we aim to explore cross-view transfer learning frameworks to increase model robustness in complex clinical scenarios. Third, we will consider incorporating type 2 fuzzy systems to further strengthen the framework's ability in handling uncertainty. From an application perspective, future work will focus on advancing online seizure prediction and further validating the transferability of the proposed framework to other EEG tasks, such as motor imagery and emotion recognition.

**Acknowledgments**

This work was supported in part by the NSFC under Grant 62176105, the Six Talent Peaks Project in Jiangsu Province under Grant XYDXX-056, Basic Research Program of Jiangsu (Grants No. BK20240315), the NSFC under Grant 62406229, and the Basic Science (Natural Science) Research Project of Higher Education Institutions in Jiangsu Province (Project No. 24KJB520039).